\documentstyle[twoside,fleqn,espcrc2]{article}
\newcommand{\AmS}{{\protect\the\textfont2
A\kern-.1667em\lower.5ex\hbox{M}\kern-.125emS}}

\begin{document}

\title{\begin{center}Transverse voltage in high-T$_{c}$ superconductors in zero
magnetic fields\end{center}}
\author{\begin{center}P. Va\v{s}ek \thanks{E-mail: vasek@fzu.cz,
FAX:(+4202) 3123184 }\\
\vspace*{3mm}
{\it{Institute of Physics ASCR, Na Slovance 2, 182 21 Prague, Czech
Republic}}\\
\vspace*{2mm}
Received January 15, 2001; revised February 15, 2001
\end{center}}
\begin{abstract}

  Longitudinal and transverse voltages have been measured on Bi - based
superconductors in zero external magnetic fields. In  close vicinity of the
superconducting transition nonzero transverse voltage has been observed
while far away  from T$_{c}$,  both above and  below  no  such voltage has been
detected. The value of the transverse resistivity  depends on the value of the
transport current.
     Several models have been discussed taking into account also the
penetration of self field due to the applied transport current. It seems that
observed results can be explained  using the  Kosterlitz - Thouless model as
a result of an   unpairing of vortex-antivortex pairs created below
T$_{KT}$  due to
fluctuations. At T$_{KT}$   free vortices and antivortices are created and can
contribute to a dissipation of energy. Their movement should also be
responsible for the observed nonzero transverse voltage.\\

\hspace{-4.5mm} $PACS$: 74.25 Fy; 74.72 Hs

\end{abstract}

\maketitle

\section{Introduction}

High Tc superconductors  represent an interesting group of substances for both
theoretical and  experimental investigations. A lot of new phenomena has been
observed for the first time on these materials. For  practical applications  the most
important properties of interest are those belonging to the group of dynamic
properties, i.e. physical effects in electric field.
From the point of view of fundamental research the well characterized single
crystalline samples are asked for. But one should have in mind that polycrystalline
ceramic materials rather than single crystals will be the main components of
superconducting devices. It is thus necessary to study  the properties of  such
polycrystalline systems as well.
  In this paper we deal with transport properties of highly textured  Bi- based (2223)
polycrystalline bulk superconductors in zero magnetic field. The temperature
dependence of longitudinal and transversal resistivity has been studied in close
vicinity of  the transition temperature.

\section{Experimental}
Two different methods  of transport measurement have been used. For the sample
in the form of thin  stripe standard six point  contact method was used with
contacts
for the measurement of longitudinal
and transverse voltage, respectively. The misalignment of transverse contacts was
corrected for by measuring the transverse voltage
in the regime where no Hall
voltage  should appear i.e. well above transition temperature where the sample was
in the normal state.


Several samples were measured by the van der Pauw method
\cite{Pauw}. In this case the
samples were either in the form of  a square or disk with the contacts equally spaced
on the perimeter of the sample. The thickness of the samples was in all cases bellow
250 $\mu$m.
The dependence of the observed nonzero transverse voltage on the current was also
measured by both   methods.
The reason for using standard six point method as a supplement to the van der Pauw
method was to verify if the non-homogeneous current distribution in the latter
method is not responsible for the observed effect.
The results for all type of the samples and methods were qualitatively the same.
Results of the van der Pauw measurement are shown in Figs. 1 and
2 for the samples of
disk shape of 15mm diameter with  thickness of the disk 120$\mu$m . From this graph
one can clearly see that in the close vicinity of the T$_c$   nonzero transverse voltage
appears which is absent both well bellow and above T$_c$. One can also see the non
symmetric shape of the transverse voltage peak.
As to the current dependence of the resistance peak its height increases with the
increasing current  passed through the sample. The current not exceeding 50 mA
was used in all measurement. For such currents we did not observe any heating
effects. This conclusion has been confirmed by two facts: no change of the
measured voltage was detected upon varying the equilibration time ( i.e. the period
between switching-on the current and measuring the voltage) and no shift was seen
in the temperature where longitudinal voltage vents effectively to zero value.


\section{Discussion}
Glazman \cite{Glaz} proposed a model for the explanation of the observed effect. Magnetic
field produced by the current going through sample can penetrate into the sample in
the form of vortices of different sign. This sign is determined by different direction
of magnetic field on the opposite edges of the sample resulting in vortices
penetration from one side and antivortices penetration from the opposite one.
Vortices and antivortices  move in opposite direction under the influence of Lorentz
force and can annihilate if the attractive interaction between them overcomes the
Lorentz force. This means that the path of vortex and antivortex will be distorted
and transverse voltage appears according to the Josephson relation. With  increasing
current the Lorentz force is stronger and the probability of annihilation decreases
i.e. for high enough current the trajectories of  vortices and antivortices are not
influenced by their interaction. They can therefore move perpendicularly to the
edges until they reach the opposite side of the sample.

According to this theory the transverse voltage value should increase for low
transport current and again decrease for high enough current. We have changed the
transport current by two orders of magnitude in our experiments and did not
observe any decrease of  transverse voltage. Similar effect was observed by
Francavilla et al. \cite{Franc} on thin sputtered YBaCuO films of unknown thickness using
currents up to 180 mA. Moreover the calculated magnetic field on the surface of our
sample is in the range of units of  $\mu$T. At such small magnetic fields the
concentration of vortices and antivortices is probably too small to initiate the
observable voltage.

In the following we will propose another explanation of  the above mentioned
effect. Since Bi based materials are strongly anisotropic their electrical behavior
should reflect the two-dimensional-like nature. Near the percolation transition when
the phase of the wave function becomes coherent along a privileged path in the
sample and  weak links among grains become irrelevant
\cite{Auslo}, the behavior should be
similar to that occurring at Kosterlitz-Thoules temperature
T$_{KT}$
\cite{Kost}. Bellow T$_{KT}$
vortex-antivortex  pairs are thermally activated . At  T$_{KT}$ these pairs spontaneously
dissociate into free vortices. These free vortices and antivortices  can move under
the influence of  external fields and cause in this way dissipation of the energy
above T$_{KT}$.
There are two possibilities how to determine T$_{KT}$.  First one, most simple, is to
determine this temperature as that at which the resistance of the sample reaches zero
value at zero magnetic field. It is commonly believed that the beginning of the
dissipation is connected with the creation of free vortices and with their  movement
\cite{Halp}.
According to the Coulomb gas model \cite{Min} there are no free vortices present below
T$_{KT}$ and no flux- flow resistance . However free vortices should , according to this
model, be generated below T$_{KT}$ provided a finite current is imposed across the
superconducting sample. The Coulomb gas model prediction for the flux-flow
resistance generated below T$_{KT}$ in this way is equivalent to a nonlinear I-V
characteristic of the form $V=I^{\alpha(T)}$. Precisely at T$_{KT}$  the prediction for the exponent
is $\alpha = 3$ which was confirmed  by many of experiments.
The exponent  $\alpha$(T) for our measurement is shown in Fig. 3. One can see that the
value $\alpha = 3$ corresponds to the temperature 104 K i.e.
T$_{KT}$  = 104 K. This value
roughly coincides with the temperature where resistance in zero magnetic field
effectively reaches zero (see Fig. 1).
It is shown in Fig. 2 that the nonzero transverse resistance starts at the same
temperature.  Set of free vortices and antivortices  created above
T$_{KT}$ is forced to
move by Lorentz and Magnus forces and dissipation can start and  according to the
Josephson relation the nonzero transverse resistance appears. This suggestion is also
supported by sudden increase of the resistivity from the side of low temperatures.


\section*{Acknowledgments}

This work has been partially supported by GACR projects Nos. 104/99/1440 and
106/99/1441 and by GAASCR under contract No. 11116.

\section*{Figure captions}

Fig.1: Temperature dependence of longitudinal resistance  R$_{xx}$
       for different current  in zero   external magnetic field

Fig.2: Temperature dependence of transverse resistance R$_{xy}$  for different
       current  in zero  external magnetic field

Fig.3: Temperature dependence of the exponent $\alpha$ for
       nonlinear I-V characteristics ($V=I^\alpha$) in zero external magnetic
       field

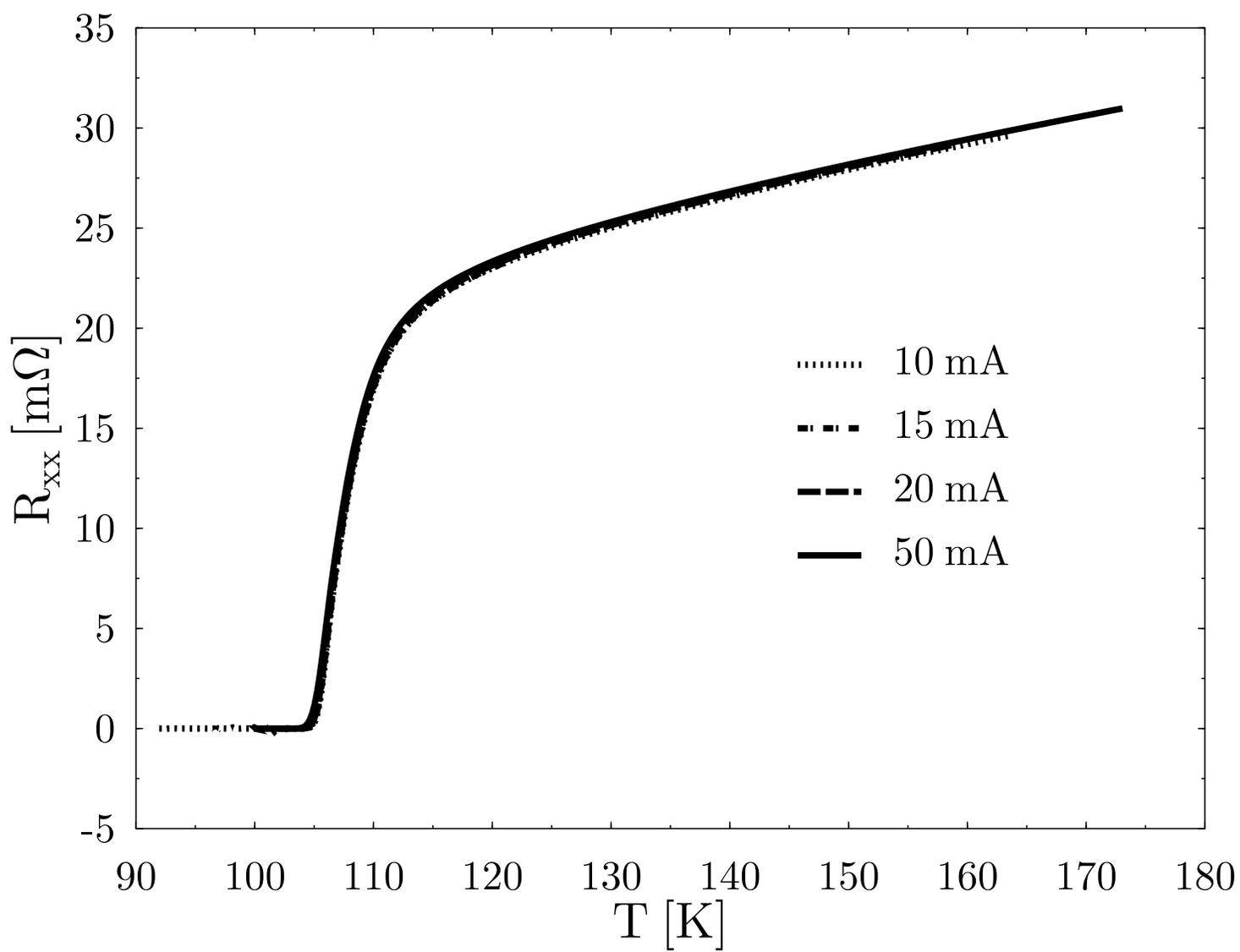
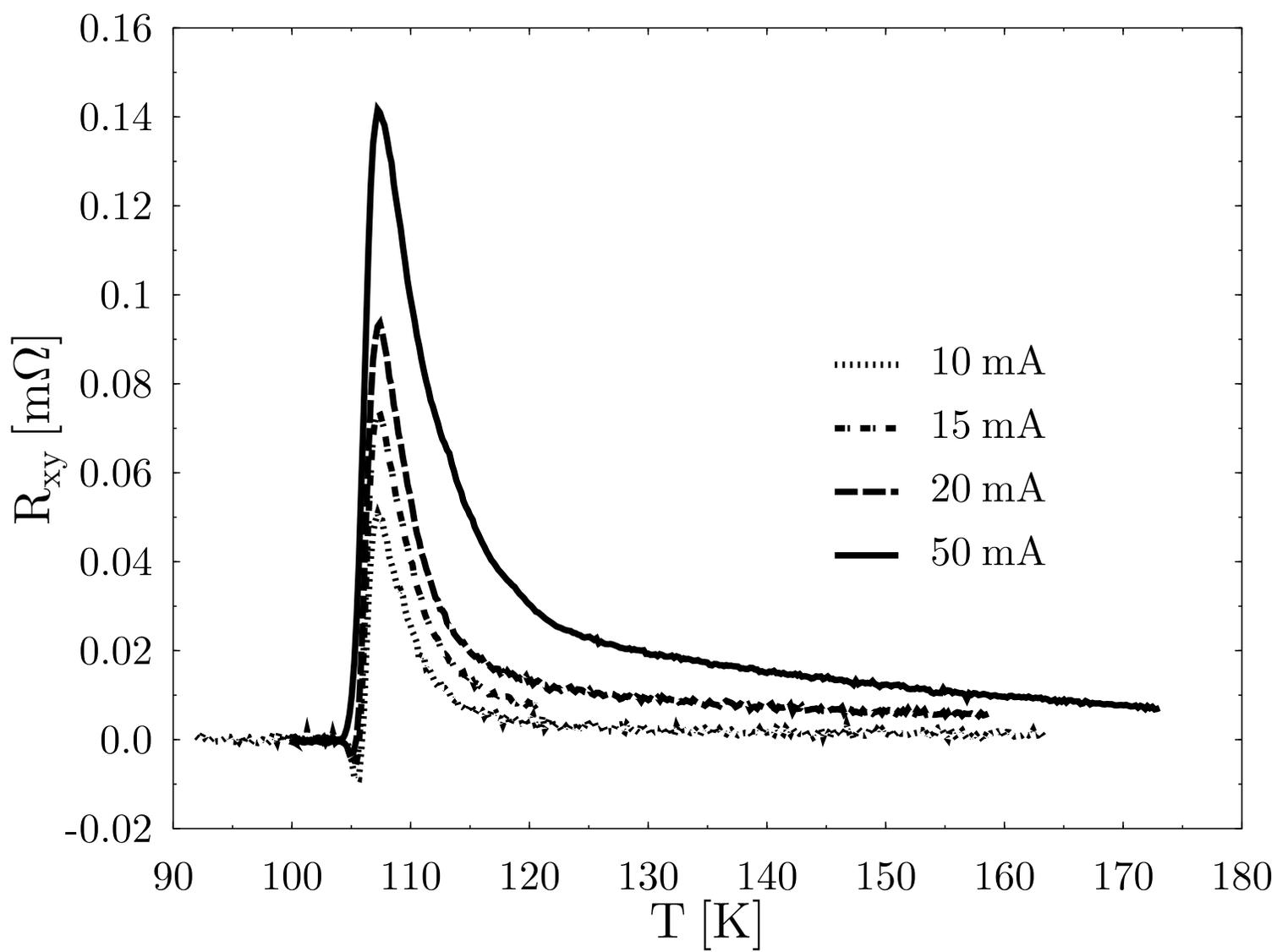
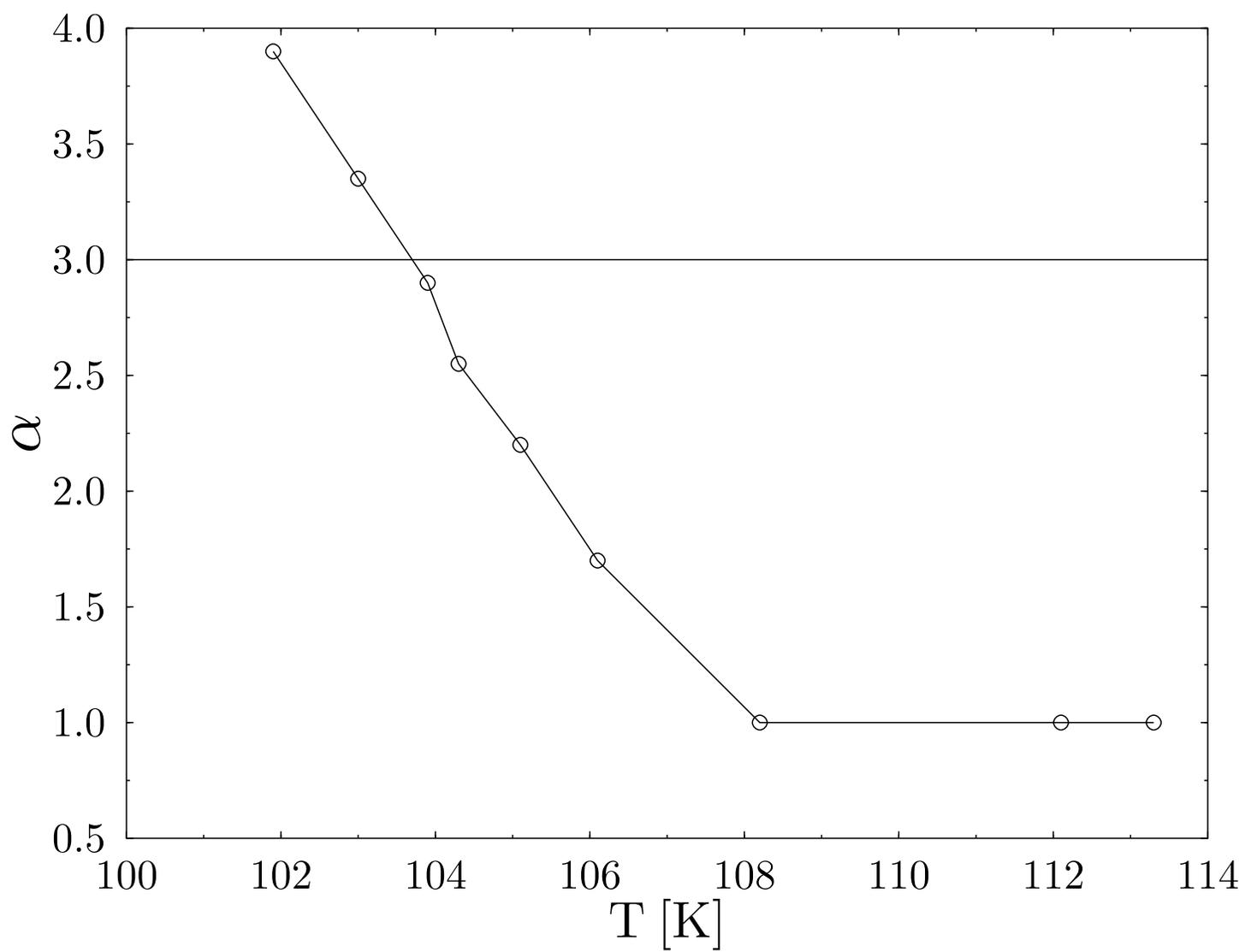
\end{document}